\documentclass[preprintnumbers,prd,aps,showpacs,amsmath,amssymb,floats]{revtex4}
\usepackage{graphicx}
\usepackage{epsfig}
\usepackage{bm}
\usepackage{amsfonts}
\usepackage{amssymb}
\usepackage{times}
\usepackage{natbib}
\usepackage[dvips]{color}

\def\lapp{\mathrel{\rlap{\raise.5ex\hbox{$<$}}
                    {\lower.5ex\hbox{$\sim$}}}}
\def\gapp{\mathrel{\rlap{\raise.5ex\hbox{$>$}}
                    {\lower.5ex\hbox{$\sim$}}}}
\begin{document}
\title{Cosmological Magnetogeneis From Extra-dimensional Gauss Bonnet Gravity}

\author{Kumar Atmjeet}
\email{katmjeet@physics.du.ac.in}
\affiliation{Department of Physics and Astrophysics, University of Delhi,
Delhi 110007, India.}
\author{Isha Pahwa}
\email{ipahwa@physics.du.ac.in}
\affiliation{Department of Physics and Astrophysics, University of Delhi,
Delhi 110007, India.}
\author{T R Seshadri}
\email{trs@physics.du.ac.in}
\affiliation{Department of Physics and Astrophysics, University of Delhi,
Delhi 110007, India.}
\author{Kandaswamy Subramanian}
\email{kandu@iucaa.ernet.in}
\affiliation{IUCAA, Post Bag 4, Ganeshkhind, Pune 411007, India.}
\date{\today}

%%%%%%%%%%%%%%%%%%%%%%%%%%%%%%%%%%%%%%%%%%%%%%%%%%%%%%%%%%%%%%%%%%%%%%%%%%%%%%%%%%%%%%%%%%%

\begin{abstract}
Generation of primordial magnetic fields during inflation 
typically
requires the breaking of conformal invariance of the Electromagnetic action.
In this paper this has been achieved naturally in a higher dimensional cosmological model with a Gauss-Bonnet term in the action. 
The evolution of the scale factor of the extra dimension (whose dynamics is influenced by the Gauss-Bonnet term) acts as the cause for
the breaking of conformal invariance. Different cases have been investigated, each of which is characterized by the number of 
higher dimensions, the value of the Gauss-Bonnet parameter, and the cosmological constant.
Many of the scenarios considered are highly constrained by the requirements that the
cosmic evolution is stable, that the normal dimensions expand and that 
there is no back reaction due to growing electric fields.
However there do exist scenarios which satisfy the above requirements 
and are well suited for magnetogenesis.
In particular, a scenario where   
the number of extra dimensions $D=4$ and
the cosmological constant is non-zero,
turns out to be best suited 
for generating primordial magnetic fields. It is shown that for these values of parameters, a scale invariant magnetic field of 
the order of $10^{-10}-10^{-9}$ $G$ can be produced. Even in these most favorable scenarios, the higher dimensional space expands 
during inflation at the same rate as the normal dimension. Hence if a mechanism could freeze the evolution of the higher dimension, 
this seems to be a viable mechanism to produce acceptable primordial magnetic fields.
\end{abstract}

%%%%%%%%%%%%%%%%%%%%%%%%%%%%%%%%%%%%%%%%%%%%%%%%%%%%%%%%%%%%%%%%%%%%%%%%%%

\maketitle
\section{Introduction}
Observations indicate the existence of 
coherent magnetic fields over all scales ranging 
from stars to galaxies and clusters of galaxies \cite{Beck12,BS05}. 
However, we still do not have a fully
satisfactory theory that explains their origin. 
There is evidence for coherent 
magnetic fields of the order
of a few $\mu$ $G$ even in galaxies at  $z \sim 1-2$ 
\cite{kpp93,kpp08,Bernet08}.
In fact, there have also been indications of a lower bound, ${B} \ge 3 \times 10^{-16}$ $G$, on Mpc scales, 
for intergalactic magnetic fields \cite{NV10,tlr11}. These $\gamma$-ray observations indicate that there 
could be an all pervasive intergalactic magnetic field filling space almost completely. 
Such a volume filling magnetic fields could most easily be explained, if they have a primordial origin.
There is, however, as yet no compelling mechanism which produces a coherent magnetic field of the required strength over 
such large scales. The existence of large scale magnetic fields indicates  that their origin could be in the early 
universe \cite{kpp93,gras01,dol03,widrow_etal12,durn13}. 
As inflation becomes the natural choice to produce coherence on large scales,
it is natural to explore the mechanism of magnetogenesis in the
context of the early universe in an inflationary scenario \cite{turn88,rat92}.
Inflation itself could have been produced by different processes. The most common
mechanism to produce an inflationary phase in the early universe involves a scalar field in a given potential. 
Different forms of the potential give different mechanism to produce inflation. Another scenario in which one 
could have an inflationary expansion of the universe is when the space time has dimensions more than $1+3$. 
Such models have been explored in literature \cite{shw83,oy86,sw87,sah89}.

In $1+3$ dimensions, the electromagnetic field action is conformally invariant \cite{parker68}. The Electric and the Magnetic field in 
such a universe decay as $1/a^{2}$. Thus in any standard cosmological model with inflation these fields will be washed out much 
before the end of inflation. In order to have magnetic fields of sufficient strength at the end of inflation, conformal invariance of electromagnetic
field action has to be broken.
\footnotetext[1]{The decay of the magnetic field can made to slow down in the case of certain open models
for the universe \cite{bty12,btyk12}. In these models the effect is purely due to geometric reasons. These are however, open models of the 
universe and hence require $K=-1$. We will not be considering such models in this paper (see also recent criticism of such models
in \cite{vs13})}
The survival of the magnetic field requires that it should decay at a slower rate with cosmological expansion 
(typically as $1/a^{\epsilon}$, where $\epsilon <<2$). Attempts have been made to generate magnetic field during inflationary period through
models based on breaking of conformal invariance of electromagnetic action 
\cite{wid02,gio04,gio08,gios08,dol93,kkt11,turn88,rat92,dur06,cdf09,ll95,ggv95,bamba04,bamba07,mukh09,bptv01,ccf08,my08,kandu10}. 
The idea of higher dimensional inflation is particularly attractive in the context of the generation of primordial magnetic fields. This approach 
could give a natural way to break conformal invariance that in turn can lead to the generation of magnetic fields.
The models of magnetogenesis based on scalar field inflation have the 
additional problem of back-reaction. The back-reaction, if present, tends to halt inflation much before the required e-fold expansion has taken place 
to solve the horizon problem in standard cosmology. This actually has given motivation  to look beyond the scalar field model of inflation 
such as higher-dimensional cosmology. Our model is based on the approach where the dynamical evolution of scale factor for extra dimensions breaks 
the conformal invariance of $1+3$ dimensional electromagnetic field action. The coupling to a dynamical scale factor of extra dimension 
could be a more natural mechanism to break conformal invariance as compared to models based on scalar field inflation which employ an arbitrary 
coupling function to implement this \cite{kkt11,kunze05,gios08}.

The plan of the paper is as follows. In the next section we have discussed the Electromagnetic action in higher dimensional models, where 
we have taken normal as well as higher dimensional subspaces to be homogeneous, isotropic and flat. Section III, focuses on  
the Gauss-Bonnet gravity and its effect on the dynamics of the universe, which is given by the solutions of the Einstein's equations. 
In section IV, we have derived the reduced $1+3$ dimensional Electromagnetic action and shown that conformal invariance of the 
Electromagnetic action is naturally broken by the dynamical scale factor of higher dimension. 
The effect of higher dimensions is embedded into the evolution equations for vector potential. We have also derived the expressions of the power 
spectrum for magnetic and electric fields in this model. In section V we have obtained the numerical solution for the evolution of scale 
factors on the basis of which we can assume exponential behaviors of the scale factors. This helps us to find an analytical solution for the vector 
potential and hence easier to compute the power spectrum of electric and magnetic fields in section VI. In this section we discuss the behavior of 
the power spectrum based on the results for different number of extra dimensions and model parameters. We have estimated the strength of
magnetic field in section VII for the various cases discussed in the preceding section. Finally we have summarized the results and possibilities 
in section VII.

We have used few notations and conventions in this work.  We have worked in natural units (i.e. $h$ $=$ $G$ $=$ $c$ $=$ $1$). We shall chose 
metric signature to be ($-,+,+,+,+....$). The Greek alphabets can take values from $0$ to $n-1$, where $n$ is the number of spatial 
dimensions involved in the theory. Lowercase Latin indices runs from $1$ to $3$ while the uppercase Latin indices takes values from $4$ to 
$3+D$, where $D$ is the number of extra dimensions in our model.
%%%%%%%%%%%%%%%%%%%%%%%%%%%%%%%%%%%%%%%%%%%%%%%%%%%%%%%%%%%%%%%%%%%%%%%%%%%%%%%%%%%%%%%%%%%%%%%%%%%%%%%%%%%%%%%%%%%%%%%%%%%%%%%%%%%%%%%%%%%%%%%%%%%%%5
\section{Electromagnetic Action in Higher Dimensional Models}
The action for the electromagnetic field in $1+3$ dimensions in a general space-time is given by,
\begin{equation}
 S_{EM}=-\int{\frac{1}{16 \pi}d^4x \sqrt{-g} F_{\mu \nu}F^{\mu \nu}},
\end{equation}
 where $F_{\mu \nu}$ is the electromagnetic field tensor given in terms of the derivatives of vector potential $A_{\mu}$, as 
 $F_{\mu \nu}=\partial_{\mu}A_{\nu}-\partial_{\nu}A_{\mu}$.
 The determinant of  the metric tensor $g_{\mu \nu}$ is denoted by $g$ .
A homogeneous and isotropic universe can be described by the line element,
\begin{equation}
 ds^2=a(\eta)^{2}(-d\eta^2+\eta_{ij}dx^{i}dx^{j}).
\end{equation}
where $a(\eta)$ is the scale factor and $\eta$ is the conformal time. Here $\eta_{ij}$ is the spatial part of Minkowski metric tensor given as,
$\eta_{\mu\nu}=diag(-1,1,1,1,....)$. Since the Electromagnetic action is conformally invariant, it can be shown that magnetic field in 
such a universe decays as $B\propto 1/a^2$. Hence if magnetic fields were to be generated in, say, an inflationary process, they would decay away
very rapidly. If we require that there should be significant magnetic field at the end of inflation, the conformal invariance of electromagnetic
action has to be broken. Only in that case one has the possibility of the dilution of magnetic field with expansion to be slower than $1/a^{2}$. 
Put alternatively, this could lead to the amplification of $a^2B$. Many different mechanisms to break conformal invariance for magnetogenesis have 
been investigated in literature
\cite{wid02,gio04,gio08,gios08,dol93,kkt11,turn88,rat92,dur06,cdf09,ll95,ggv95,bamba04,bamba07,mukh09,bptv01,ccf08,my08,kandu10}. 
One of the mechanisms used is to make the coefficient 
of $F^{\mu \nu}F_{\mu \nu}$ a time dependent function instead of it being a constant as in the standard electrodynamics. In the following section, 
we have explored the possibility of extra dimensional models with Gauss-Bonnet term provided such a function for breaking of conformal invariance 
in the reduced four dimensional action. We consider a higher dimensional space-time which has $D$ extra spatial dimensions in addition to the normal 
$1+3$ dimensions.
We further assume that the spatial part of normal as well as extra dimensional subspace is homogeneous, isotropic and flat.
The line-element for such a universe can be given by,
\begin{equation}
 ds^2=\tilde{g}_{\mu\nu}dx^{\mu}dx^{\nu}=-dt^2+a^2(t)\eta_{ij}dx^{i}dx^{j}+b^2(t) \eta_{IJ}dx^{I}dx^{J}
\label{metric}
\end{equation}
where $\tilde{g}_{\mu\nu}$ is the metric tensor for higher dimensional space-time. The functions $a(t)$ and $b(t)$ are the scale factors of normal 
and extra dimensions, respectively. The action for the electromagnetic field in higher dimensional theory is taken to be,
\begin{equation}
 S=\frac{-1}{16 \pi}\int{d^{4+D}x\sqrt{-\tilde{g}}\tilde{F}_{\mu \nu}\tilde{F}^{\mu \nu}}
\end{equation}
Here $\tilde{g}$ is the determinant of the higher dimensional metric tensor 
$\tilde{g}_{\mu \nu}$, and $\tilde{F}_{\mu \nu}=\partial_{\mu}\tilde{A}_{\nu}-\partial_{\nu}\tilde{A}_{\mu}$ 
is similarly the higher dimensional electromagnetic field tensor in terms of higher dimensional vector field $\tilde{A}_{\mu}$. As we will 
see later, conformal invariance for $1+3$ dimensional electromagnetic action is broken by the presence of the dynamical 
scale factor ($b(t)$) of higher dimensions in this scenario \cite{kkt11,kunze05,gios08}.

%%%%%%%%%%%%%%%%%%%%%%%%%%%%%%%%%%%%%%%%%%%%%%%%%%%%%%%%%%%%%%%%%%%%%%%%%%%%%%%%%%%%%%%%%%%%%%%%%%

\section{Gauss-Bonnet Gravity}
Higher dimensional gravity is a natural generalization of $1+3$ dimensional gravity. The literature is abundant with higher dimensional 
models like String theory \cite{bd85}, Brane-World, etc. \cite{rs99,rsp99,add98,vsm07,mt04} to name a few. 
In this paper we study in particular, Gauss-Bonnet gravity
in higher-dimensional scenario \cite{love71,dad05,mad86}. The standard invariant Einstein-Hilbert action leads to equations of motion for the 
geometry (Einstein's equations) which are second order in the components of the metric tensor. Since the dynamics of the universe 
today is well explained within the context of $1+3$ dimensional Einstein-Hilbert action, the effect of considering a higher dimensional 
action should be such that we recover the standard $1+3$ dimensional Einstein equations today, at least in a low energy approximation.
A natural generalization to higher dimensional action will involve the Gauss-Bonnet term. In $1+3$ dimensions, for space without 
boundary (as in case of cosmology) the Gauss-Bonnet term becomes the total divergence. Thus, it does not contribute to the equations 
of motions in $1+3$ dimensions. This term, however, gives a non-zero contribution in higher dimensions and hence it is interesting to 
look for its effect in higher dimensional cosmology. This term becomes very small today as
it varies as square of the curvature.

We consider, the dynamics of the universe governed by the action of the form \cite{abm07,chto12},
\begin{equation}
 S \, = \, \int d^{4+D}x \sqrt{-\tilde{g}} \, (\mathcal{L}_{matter}+\tilde{\mathcal{L}}_{EM}-\frac{M^{D+2}}{2}(\tilde{R}\, + \chi \, \tilde{G}) ) \label{action}
\end{equation}
where $\tilde{R}$ is the `$1+3+D$' dimensional Ricci scalar and $\chi$ is the Gauss-Bonnet parameter.
$M$ is the higher dimensional Planck mass which is related to 
$1+3$ dimensional Planck mass as $M_{pl}^2=b^{D}M^{D+2}$.
We may define a parameter $\theta$ as $\theta = \chi M^{2}$ which sets the scale of the theory.
The Gauss-Bonnet term $\tilde{G}$ is given by,
\begin{equation}
 \tilde{G}=\tilde{R}^2-4\tilde{R}_{\mu \nu}\tilde{R}^{\mu \nu}+\tilde{R}_{\mu \nu \lambda \sigma}\tilde{R}^{\mu \nu \lambda \sigma}
\end{equation}
and $\tilde{\mathcal{L}}_{EM}$ is Lagrangian density of the electromagnetic field given by,  
$\tilde{\mathcal{L}}_{EM}=-\frac{1}{16 \pi}\tilde{F}_{\mu \nu}\tilde{F}^{\mu \nu}$ in the higher dimensions.

The Electromagnetic filed is assumed to be a test field and hence does not contribute to the evolution of the background.
The source term that affects the evolution of geometry is the energy-momentum tensor arising from $\mathcal{L}_{matter}$ of the form,
\begin{displaymath}
 T^{\mu}_{\nu}=(-\rho,P_{1},P_{1},P_{1},P_{2},P_{2},...)
\end{displaymath}
where $\rho$ is the energy density. $P_{1}$ and $P_{2}$ are isotropic pressure in normal and extra dimensions respectively.
We adopt $P_1 = w_1 \rho$, $P_2=w_2\rho$, 
where $w_1$ and $w_2$ are the equation of state parameters for the constituent present in normal and extra dimensions, respectively.

The continuity equation in higher dimension ($T^{\mu\nu}_{;\nu}=0$) implies,
\begin{equation}
 \rho(t)=\frac{\rho_{0m}a_0^{(3(1 + w_1))}b_0^{(D(1 + w_2))}}{a(t)^{(3(1 + w_1))}b(t)^{(D(1 + w_2))}}
\end{equation}
The Einstein tensor $G_{\mu}^{\nu}$ for the line element given in Eq.~(\ref{metric}) is derived for 
this action assuming that electromagnetic field is a test field. The corresponding non zero spatial components within $3$-space and within extra spatial 
dimensions are equal \cite{pahwa08}.
\begin{eqnarray}
 -\frac{\rho}{M^{D+2}} &=& -3  \frac{\dot{a}^2}{a^2} - 3  D\frac{\dot{a}}{a}\frac{\dot{b}}{b}- \frac{D(D-1)}{2} \frac{\dot{b}^2}{b^2} + 12 D \chi \frac{\dot{a}^3 \dot{b}}{a^3 b} +18 D(D-1) \chi \frac{\dot{a}^2 \dot{b}^2}{a^2 b^2}  \nonumber\\
       & & + 6 D(D-1)(D-2)\chi \frac{\dot{a} \dot{b}^3}{a b^3}+\frac{D(D-1)(D-2)(D-3)}{2} \chi \frac{\dot{b}^4}{b^4}  \label{Eeq00}\\
\frac{P_1}{M^{D+2}}  &=& -2 \frac{\ddot{a}}{a}-D \frac{\ddot{b}}{b}-2D \frac{\dot{a}}{a}\frac{\dot{b}}{b} - \frac{\dot{a}^2}{a^2} - \frac{D(D-1)}{2} \frac{\dot{b}^2}{b^2} +8D\chi \frac{\ddot{a}\dot{a}\dot{b}}{a^2b}+4D(D-1)\chi \frac{\dot{b}^2\ddot{a}}{b^2a}+4D\chi \frac{\dot{a}^2\ddot{b}}{a^2b} \notag \\
       & & +2D(D-1)(D-2)\chi \frac{\dot{b}^2\ddot{b}}{b^3}+6D(D-1)\chi \frac{\dot{a}^2\dot{b}^2}{a^2b^2}+4D(D-1)(D-2)\chi \frac{\dot{b}^3\dot{a}}{b^3a}\notag \\
       & & +D(D-1)(D-2)(D-3)\chi \frac{\dot{b}^4}{2b^4} +8D(D-1)\chi  \frac{\ddot{b}\dot{b}\dot{a}}{b^2a}\hspace{2cm} \forall \ i \label{Eeqii}\\
\frac{P_2}{M^{D+2}}  &=& -(D-1) \frac{\ddot{b}}{b}-3 \frac{\ddot{a}}{a}-3(D-1) \frac{\dot{a}}{a}\frac{\dot{b}}{b}-(D-1) \left(\frac{D}{2}-1\right) \frac{\dot{b}^2}{b^2} -3 \frac{\dot{a}^2}{a^2}+24(D-1)\chi \frac{\ddot{a}\dot{a}\dot{b}}{a^{2}b} \notag \\
       & & +6(D-2)(D-1)\chi \frac{\dot{b}^2\ddot{a}}{b^{2}a}+12(D-1)\chi \frac{\dot{a}^2\ddot{b}}{a^2b}+2(D-1)(D-2)(D-3)\chi \frac{\dot{b}^2\ddot{b}}{b^3} \notag \\
       & & +6(D-1)(D-2)(D-3)\chi \frac{\dot{b}^3\dot{a}}{b^3a}+(D-4)(D-1)(D-2)(D-3)\chi \frac{\dot{b}^4}{2b^4} +12\chi \frac{\dot{a}^2\ddot{a}}{a^3}\notag \\
       & & +18(D-1)(D-2)\chi \frac{\dot{a}^2\dot{b}^2}{a^2b^2}+12(D-2)(D-1)\chi \frac{\ddot{b}\dot{b}\dot{a}}{b^{2}a}+12(D-1)\chi \frac{\dot{a}^3\dot{b}}{a^3b} \hspace{1cm} \forall \ I \label{EeqII}
\end{eqnarray}
Augmented by the equation of state, the solution of these equations gives the evolution of $a(t)$ and $b(t)$.

%%%%%%%%%%%%%%%%%%%%%%%%%%%%%%%%%%%%%%%%%%%%%%%%%%%%%%%%%%%%%%%%%%%%%%%%%%%%%%%%%%%%%%%%%%%%%%%%%%%%%%%%%%%%%%%%%%%%%%%

\section{Magnetogenesis in Gauss-Bonnet Gravity}
We consider, the universe to be described by the metric given in Eq.~(\ref{metric}). By performing dimensional reduction we get a 1+3- dimensional 
effective Electromagnetic action as,
\begin{equation}
 S_{em}=\int{d^{4}x\sqrt{-g}\mathcal{L}_{EM}\left(\frac{b}{b_{0}}\right)^{D}},
 \label{redaction}
\end{equation}
where, $\mathcal{L}_{EM}=b_{0}^{D}\Omega_{D}\tilde{\mathcal{L}}_{EM}$. $\Omega_{D}$ is the co-ordinate volume of extra dimensions which is
assumed to be finite and $g$ is the determinant of $1+3$ -dimensional metric tensor $g_{\mu\nu}$. $\mathcal{L}_{EM}$ 
is the equivalent $1+3$ -dimensional Lagrangian density for $1+3$-dimensional vector potential $A_{\mu}$ (for $\mu= 0$ to $3$) defined by,
\begin{equation}
 \mathcal{L}_{EM}=\frac{-1}{16\pi}F_{\mu\nu}F^{\mu\nu}
\end{equation}
The reduced $1+3$-dimensional electromagnetic action corresponds to a $1+3$- dimensional vector potential given by,
\begin{equation}
 A_{i}=(\Omega_{D}b_{0}^{D})^{1/2}\tilde{A}_{i}
\end{equation}
We have considered a simple case where the nature of $A^{\mu}$ is assumed to be such that it depends on only $1+3$ -dimensional co-ordinates. 
This ensures that the derivatives of $A^{\mu}$ w.r.t. extra dimensional co-ordinates vanish. Further we assume that components of
$A^{\mu}$ for $\mu\geqslant4$ are zero. 
The choice of field configuration (i.e. the field is completely confined to external space) allows one to do such reduction and the reduced field is 
identified as the $1+3$-dimensional vector potential in our case. Another approach followed in \cite {gszhuk04} leads to terms containing scalar fields as well. 
This differs from an approach of considering five dimensional field with electromagnetic components confined in $1+3$-dimensions \cite{memfb12}.
The action for electromagnetic field given by Eq.~(\ref{redaction}) is no more conformally invariant.
The presence of dynamical extra-dimensional scale factor of higher dimension breaks the conformal invariance of the $1+3$ -dimensional electromagnetic action.
Maxwell's equations for electromagnetic fields is obtained by varying the action with respect to the reduced four dimensional vector potential. 
\begin{equation}
 \frac{1}{\sqrt{-{g}}}\partial_{\mu}\left[\sqrt{-g}\left(\frac{b}{b_{0}}\right)^{D}{F}^{\mu\nu}\right]=0 \label{maxwell}
\end{equation}
We work in the radiation gauge i.e. $A_{0}=0$, $\partial_{i}A^{i}=0$.
Maxwell's equation, Eq.~(\ref{maxwell}) then takes the form,
\begin{equation}
 \ddot{A_{i}}+\left(\frac{\dot{a}}{a}+D\frac{\dot{b}}{b}\right)\dot{A_{i}}-\frac{\partial_{j}\partial_{j}}{a^2}A_{i}=0
\end{equation}
where dot is the derivative with respect to time, $t$.
It is convenient to work in terms of conformal time co-ordinate $\eta$, defined as,
\begin{equation}
 \eta=\int{\frac{dt}{a(t)}}   \label{etaref}
\end{equation}
Re-expressing Eq ($16$) in terms of $\eta$, we get
\begin{equation}
 A_{i}''(\eta,x)+D\frac{b'}{b}A_{i}'(\eta,x)-\partial_{j}\partial_{j}A_{i}(\eta,x)=0
\end{equation}
where prime is the derivative with respect to $\eta$.
It can be seen that the presence of dynamical extra dimensional scale factor breaks the conformal invariance of 
Electromagnetic action in $1+3$-dimensions and may work to amplify the field. It is this feature of the extra dimension 
which we exploit here for our purpose.
\subsection{Evolution of Normal Modes}
It will be useful to describe the evolution of vector potentials in terms of their Fourier modes,given by
\begin{equation}
 A_{l}(x,t)=\sqrt{4 \pi}\int{\frac{d^3k}{(2\pi)^3}\sum_{\lambda=1}^2 \epsilon_{\lambda l}(k)\left[b_{\lambda}(k)A(k,\eta) e^{ikx}+b_{\lambda}^{\dag}(k)A^{*}(k,\eta)e^{-ikx}\right]},
\end{equation}
where, $k$ represents the wave number of momentum modes of vector potential. The polarization vectors $\epsilon_{\lambda i}$ are defined as
\begin{equation}
 \epsilon_{0}^{\mu}=\left(\frac{1}{a},0\right) \hspace{1cm} \epsilon_{\lambda}^{\mu}=\left(0,\frac{\tilde{\epsilon}_{\lambda}^{i}}{a}\right), \hspace{1cm} \epsilon_{3}^{\mu}=\left(0,\frac{1}{a}\frac{k^{i}}{k}\right),
\end{equation}
and $\lambda$, corresponds to two orthonormal transverse polarizations \cite{my08,kandu10}. 
The 3-vectors $\tilde{\epsilon}_{\lambda}^{i}$ are unit vectors orthogonal to ${\bf k}$ and to each other.
The conjugate momentum $\varPi^{i}(\eta,x)$
 for the vector potential is given by,
\begin{equation}
 \varPi^{i}(\eta,x)=\frac{\delta S}{\delta A_{i}'}=\left(\frac{b}{b_{0}}\right)^{D}a^{2}(\eta)g^{ij}A_{j}'(\eta,x).
\end{equation}
Here $b_{\lambda}(k)$ and $b_{\lambda}^{\dag}(k)$ are the annihilation and creation operators 
(not to be confused with the scale factor $b(t)$), 
which satisfy the relations,
\begin{equation}
 [b_{\lambda}(k),b_{\lambda'}^{\dag}(k)]=\delta_{\lambda,\lambda'}\delta^3(k-k'), \hspace{1cm} [b_{\lambda}(k),b_{\lambda'}(k)]=[b_{\lambda}^{\dag}(k),b_{\lambda'}^{\dag}(k)]=0
\end{equation}
The form of the polarization vector ensures that the coulomb gauge conditions as well as the quantization conditions are satisfied. i.e.
\begin{equation}
 [A_{l}(t,x),\varPi^{j}(t,y)]= i\int{\frac{d^{3}k}{(2\pi)^3}e^{i k(x-y)}\left(\delta_{l}^{j}-\delta_{lm}\frac{k^{j}k^{m}}{k^2}\right)}
 \label{Pol}
\end{equation}

The Fourier co-efficients,  $\bar{A}(k,\eta)=a A(k,\eta)$ satisfy the equation
\begin{equation}
\bar{A}''(k,\eta)+D\frac{b'}{b}\bar{A}'(k,\eta)+k^2\bar{A}(k,\eta)=0.
\end{equation}
Defining a new variable $\mathcal{A}(k,\eta)$ by $\mathcal{A}(k,\eta)=(b/b_{0})^{D/2} \bar{A}(k,\eta)$ and expressing the differential 
equation in terms of the new variable, we get
\begin{equation}
  \mathcal{A}''(k,\eta)+\left[k^2-\frac{D}{2}\frac{b''}{b}-\frac{D}{2}\left(\frac{D}{2}-1\right)\frac{b'^{2}}{b^2}\right]\mathcal{A}(k,\eta)=0.
  \label{VP}
\end{equation}
This can be expressed in a compact form as
\begin{equation}
 \mathcal{A}''(k,\eta)+\left[k^2-V(\eta)\right]\mathcal{A}(k,\eta)=0,
\end{equation}
where,
\begin{equation}
 V(\eta)=\frac{D}{2}\frac{b''}{b}+\frac{D}{2}\left(\frac{D}{2}-1\right)\frac{b'^{2}}{b^2}.
\end{equation}
\subsection{Power Spectrum}
The 1+3-dimensional Energy-Momentum Tensor for electromagnetic field described by the action in Eq.~(\ref{redaction}) is given by,
\begin{equation}
 T_{\mu \nu}^{EM}=\frac{-2}{\sqrt{-g}}\left(\frac{b}{b_{0}}\right)^{D}\frac{\partial(\sqrt{-g}\mathcal{L}_{EM})}{\partial{g}^{\mu\nu}}
 \label{EMtensor}
\end{equation}
Expressing $\mathcal{L}_{EM}$ in terms of electromagnetic field tensor $F_{\mu\nu}$ and metric tensor $g_{\mu\nu}$, Eq.~(\ref{EMtensor}) takes the form,
\begin{equation}
  T_{\mu \nu}^{EM}=\frac{1}{4\pi}\left(\frac{b}{b_{0}}\right)^{D}\left[g^{\delta\gamma}F_{\mu\gamma}F_{\nu\delta}-\frac{1}{4}g_{\mu\nu}F_{\alpha\beta}F^{\alpha \beta}\right]
\end{equation}
The electric field component($T_{0}^{0(E)}$) and magnetic field component ($T_{0}^{0(B)} $) of the Energy-Momentum tensor is obtained as,
\begin{equation}
 T_{0}^{0(B)}=-\frac{1}{16 \pi}\left(\frac{b}{b_{0}}\right)^{D}g^{ij}g^{ml}F_{im}F_{jl},\hspace{5mm} T_{0}^{0(E)}=-\frac{1}{8 \pi}\left(\frac{b}{b_{0}}\right)^{D}\frac{g^{ij}}{a^{2}}A_{i}'A_{j}'
\end{equation}
Vacuum expectation value of energy density for magnetic field ($\rho_B$) contribution is given by,
\begin{equation}
 \rho_B=\left<0\left|-T_{0}^{0B}\right|0\right>=\frac{1}{16\pi}\left(\frac{b}{b_{0}}\right)^{D}\left<0\left|(\partial_{i}A_{m}-\partial_{m}A_{i})(\partial_{j}A_{l}-\partial_{l}A_{j})g^{ij}g^{ml}\right|0\right>
 \label{MED}
\end{equation}
Using the Fourier transformation defined in previous section, Eq.~(\ref{MED}) is rewritten as
\begin{eqnarray}
\rho_{B}&=&\frac{1}{16\pi}\left(\frac{b}{b_{0}}\right)^{D}4\pi \int{\frac{d^3k}{(2\pi)^3}g^{ij}g^{ml}\sum_{\lambda=1}^{2}\left(\epsilon_{\lambda m}(k)k_{i}-\epsilon_{\lambda i}(k)k_{m}\right)\left(\epsilon_{\lambda l}(k)k_{j}-\epsilon_{\lambda j}(k)k_{l}\right)\left|A(k,\eta)\right|^2} \nonumber\\
 &=&\frac{1}{2}\left(\frac{b}{b_{0}}\right)^{D} \int{\frac{d^3k}{(2\pi)^3}\frac{k^2}{a^2}\left[\sum_{\lambda=1}^{2}\epsilon_{\lambda}^{l}(k)\epsilon_{\lambda l}(k)-\frac{k^{l}k_{j}}{k^2}\sum_{\lambda=1}^{2}\epsilon_{\lambda}^{j}(k)\epsilon_{\lambda l}(k)\right]\left|A(k,\eta)\right|^2}\nonumber \\
 &=&\frac{1}{2\pi^2}\left(\frac{b}{b_{0}}\right)^D\int{dk\frac{k^4}{a^2}\left|A(k,\eta)\right|^{2}}
\end{eqnarray}

Here we have used the relations \cite{my08,kandu10},
\begin{eqnarray}
 \left<0\left|b_{\lambda}(k)b_{\lambda'}^{\dag}(k')\right|0\right>=(2\pi)^3\delta^{3}(k-k')\delta_{\lambda\lambda'},   \nonumber \\
 \left<0\left|b_{\lambda}(k)b_{\lambda'}(k')\right|0\right>=\left<0\left|b_{\lambda}^{\dag}(k)b_{\lambda'}^{\dag}(k')\right|0\right>=0
\end{eqnarray}
\begin{equation}
 \sum_{\lambda=1}^{2}\epsilon_{\lambda}^{i}(k)\epsilon_{\lambda j}(k)=\delta_{j}^{i}-\delta_{jl}\frac{k^i k^l}{k^2}
\end{equation}
In terms of redefined variable $\mathcal{A}(k,\eta)$, we have,
\begin{equation}
 \rho_{B}=\frac{1}{2\pi^2}\int{\left(\frac{k}{a}\right)^{4}\left|\mathcal{A}(k,\eta)\right|^{2}dk}
\end{equation}
Hence, the power spectrum corresponding to magnetic field $\left(\dfrac{d\rho_{B}}{dlnk}\right)$ is be given by,
\begin{equation}
 \frac{d\rho_{B}}{dlnk}=\frac{1}{(2 \pi)^2}k(\frac{k}{a})^4\left|\mathcal{A}(k,\eta)\right|^2.
 \label{PB}
\end{equation}
Similarly for Electric field, the power spectrum $\left(\dfrac{d\rho_{E}}{d lnk}\right)$ can be expressed as
\begin{equation}
 \frac{d\rho_{E}}{d lnk}=\frac{b^{D}}{2\pi^2}\frac{k^3}{a^4}\left|\left(\frac{\mathcal{A}(k,\eta)}{b^{D/2}}\right)'\right|^2
 \label{PE}
\end{equation}
Here, $\rho_{E}$ is the vacuum energy density contribution of electric field. The expressions for power spectrum of magnetic and 
electric fields have forms similar to that obtained in references \cite{my08,kandu10}. In Eq.~(\ref{PB}) \& ~(\ref{PE}). The factor $b^{D/2}$ appears
in the same way as time dependent coupling function $f(\phi)$ in these references. The coupling function $f(\phi)$ is being used to 
break the conformal invariance of electromagnetic action in these references, whereas the scale factor for extra dimensions
$b(t)$ does the job in our model. In this sense, the mechanism of breaking conformal invariance emerges more naturally in our case. In order to make 
numerical estimates of the power-spectrum, we need to consider specific 
models for the evolution of the scale factors.

%%%%%%%%%%%%%%%%%%%%%%%%%%%%%%%%%%%%%%%%%%%%%%%%%%%%%%%%%%%%%%%%%%%%%%%%%%%%%%%%%%%%%%%%%%%%%%%%%%%%%%%%%%%%%%%%%%%%%%%

\section{Numerical solutions for scale factors} \label{numerical}
The evolution of the two scale factors is governed by the Einstein's Eqs.~(\ref{Eeq00})-~(\ref{EeqII}).
These solutions have been discussed in detail in the reference \cite{pahwa08} whose results we adopt here for further calculations.
As has been discussed in
\cite{pahwa08},
there exist both
stable and unstable solutions for the scale factors.
Since the unstable solutions are not of physical interest in the context
of our work, we concentrate
only to the stable solutions here.
These equations are solved for $D$ number of extra dimensions and the redefined Gauss-Bonnet parameter $\theta$.
Further, as pointed out in reference \cite{pahwa08}, unless we have some very special situations, the energy density in the 
Universe decreases rapidly with time. Hence, we begin by considering the vacuum case when $T^{i}_{k}=0$.
The initial condition for ${\dot{b}}/{b}$ is also a parameter in the theory whereas the initial 
condition for ${\dot{a}}/{a}$ can be calculated in terms of ${\dot{b}}/{b}$ from Eq.~(\ref{Eeq00}). 
The solutions found in \cite{pahwa08} suggest that we can choose the form for $a(t)$ and $b(t)$
asymptotically to be,
\begin{equation}
 a(t)\propto e^{\alpha t}, \hspace{2cm}  b(t)\propto e^{\beta t}
\label{asssoln}
\end{equation}
The values of these exponents $\alpha$ and $\beta$ 
are given in Table~(\ref{table1}) 
for different number of extra dimensions, 
$D\geq2$. 
It is to be noted that the sign of these exponents $\alpha$ 
and $\beta$ determines whether 
inflation or contraction obtains in the respective dimensions. 
\begin{center}
\begin{table}[!htpb]
\begin{tabular}{|c|c||c|c|c|c||c|c|c|c|}
\hline
\multicolumn{2}{|c|}{} & \multicolumn{4}{|c|}{Stable Solutions}&\multicolumn{4}{|c|}{Unstable Solutions}\\
\hline
$D$ & $\theta=\chi M^2$ & $\alpha$ & $\beta$ & $\xi = -\frac{D}{2}\frac{\beta}{\alpha}$ &$n_{B}=4+2n$& $\alpha$ & $\beta$ & $\xi = -\frac{D}{2}\frac{\beta}{\alpha}$&$n_{B}=4+2n$\\
\vspace{-0.2cm}
&&&&&&&&&\\
\hline \hline
$2$ & $-1.0$ &$0.750$ & $-0.541$ & $0.722$&$4.556$ & $-0.750$ & $0.541$ & $0.722$&$4.556$\\
\hline
   &$-0.5$&$1.060$&$-0.766$&$0.722$&$4.556$&$-1.060$&$0.766$&$0.722$&$4.556$\\
 \hline \hline
$3$&$-1.0$&$0.809$&$-0.309$&$0.573$&$4.854$&$-0.809$&$0.309$&$0.573$&$4.854$\\
   &      &$-0.309$&$0.809$&$3.927$&$-1.854$&$0.309$&$-0.809$&$3.927$&$-1.854$\\ 
   \hline
   &$-0.5$&$1.144$&$-0.437$&$0.573$&$4.854$&$-1.144$&$0.437$&$0.573$&$4.854$\\
   &      &$-0.437$&$1.144$&$3.927$&$-1.854$&$0.437$&$-1.144$&$3.927$&$-1.854$\\
   \hline
   &$-0.1$&$2.558$&$-0.977$&$0.573$ &$4.854$& $-2.558$&$0.977$&$0.573$&$4.854$\\
   &      &$-0.977$&$2.558$&$3.927$&$-1.854$&$0.977$&$-2.558$&$3.927$&$-1.854$\\
\hline \hline
$4$&$-1.0$& $0.846$& $-0.218$ &$0.516$&$4.968$& $-0.846$ & $0.218$ &$0.516$&$4.968$\\
   &      & $-0.390$ & $0.525$ & $2.691$&$0.617$& $0.390$ & $-0.525$ & $2.691$&$0.617$\\
\hline
   & $-0.5$ & $1.197$ & $-0.309$ &$0.516$&$4.968$& $-1.197$ & $0.309$ &$0.516$&$4.968$\\
   &        & $-0.552$& $0.742$ & $2.691$&$0.617$ & $0.552$& $-0.742$ & $2.691$&$0.617$\\
 \hline
   &$-0.1$&$2.677$&$-0.691$&$0.516$&$4.968$&$-2.677$&$0.691$&$0.516$&$4.968$\\
   &      &$-1.234$&$1.661$&$2.691$&$0.617$&$1.234$&$-1.661$&$2.691$&$0.617$\\
 \hline
 \hline
$10$ &$-1.0$&$0.918$&$-0.080$&$0.436$&$4.872$&$-0.918$&$0.080$&$0.436$&$4.872$\\
    &      &$-0.588$&$0.218$&$1.855$&$2.288$&$0.588$&$-0.218$&$1.855$&$2.288$\\  
\hline 
  &$-0.5$&$1.298$&$-0.113$&$0.436$&$4.872$&$-1.298$&$0.113$&$0.436$&$4.872$\\
  &      &$-0.832$&$0.308$&$1.855$&$2.288$&$0.832$&$-0.308$&$1.855$&$2.228$\\
  \hline 
  &$-0.1$&$2.903$&$-0.253$&$0.436$&$4.872$&$-2.903$&$0.253$&$0.436$&$4.872$\\
  &      &$-1.860$&$0.690$&$1.855$&$2.288$&$1.860$&$-0.690$&$1.855$&$2.288$\\
 \hline
\end{tabular}
\caption{Asymptotic solutions for different number of extra dimensions}
\label{table1}
\end{table}
\end{center}

There are some key features of the solutions from Table~(\ref{table1}) which are worth highlighting. 
As we mentioned above, 
all the solutions show asymptotically exponential behavior for the scale factors of normal as well as extra dimensions.
The evolution of scale factors $a(t)$ and $b(t)$ are opposite in nature, which means if one scale factor inflates the other contracts or vice-versa.
The exponents $\alpha$ and $\beta$ depend on $\theta$ (Gauss Bonnet parameter) but their ratios are found to be independent
of $\theta$, for any given $D$ and sign of $\alpha$.
For the magnetic and electric field power spectrum calculations, as we will see in the next section, it is the ratio of the exponents
which  matters. This implies that the the choice of Gauss-Bonnet parameter only affects the evolution of scale factors but not the nature of
magnetic  field power spectrum. The details of these solutions will be used later to calculate the magnetic field spectrum and 
strength. We will see in the next section that an inflationary solution for normal dimensions and contraction for extra dimensions 
does not lead scale invariant spectrum for magnetic field.
We have also therefore investigated solutions including a cosmological constant type term ($\bar{\lambda}$). 
This works as a parameter in our theory. In this case, we first get vacuum type solutions, where the normal space inflates and the extra dimensional 
space simultaneously contracts (or vice versa). In addition, we also get a new type of stable solution, whereby
both the normal as well as extra dimension are inflating. The situation in which both the dimensions are inflating or deflating simultaneously, 
and $D=4$, turns out to be very interesting as it gives scale invariant spectrum for magnetic field (described in the next section).
A list of all the asymptotic solutions for $D=4$, $\theta=0.1$ is given in Table~(\ref{table2}).

\begin{center}
\begin{table}[!htpb]
\begin{tabular}{|c||c|c|c|c||c|c|c|c|c|}
\hline
\multicolumn{1}{|c|}{} & \multicolumn{4}{|c|}{Stable Solutions}&\multicolumn{4}{|c|}{Unstable Solutions}\\
\hline
$\bar{\lambda}$ & $\alpha$ & $\beta$ & $\xi = -\frac{D}{2}\frac{\beta}{\alpha}$ &$n_{B}=4+2n$& $\alpha$ & $\beta$ & $\xi = -\frac{D}{2}\frac{\beta}{\alpha}$&$n_{B}=4+2n$\\
\vspace{-0.2cm}
& & & & & & & &\\
\hline \hline
$0.000$& $0.846$& $-0.218$ &$0.516$&$4.968$& $-0.846$ & $0.218$ &$0.515$&$4.968$\\
     & $-0.390$ &$0.525$ &$2.691$&$0.617$& $0.390$ & $-0.525$ &$2.691$&$0.617$\\
\hline 
$0.001$ & $-0.390$  & $-0.524$   &  $ 2.690$    & $ 0.625$ &$0.390$  & $0.524$    & $2.690$  & $0.625$\\ 
        & $0.846$  & $-0.218$   &    $0.515$    &  $4.969$ &$-0.846$  & $0.218$    & $0.510$   & $4.969$\\
        & $-0.018$& $0.000$       &     $0.000$    & $4.000$  & $0.018$ & $0.000$       & $0.000$    & $4.000$\\
        & $0.007$& $0.007$  &  $-2.000$     &   $0.000$&$-0.007$  & $-0.007$   & $-2.000$  & $0.000$\\
\hline 
$0.050$ & $0.827$& $-0.216$&$0.532$&$4.935$ & $-0.827$ & $0.216$ &$0.532$&$4.935$\\
        & $-0.384$ & $0.505$ &$2.630$&$0.740$ & $0.384$ & $-0.505$ &$2.630$&$0.740$\\
        & $0.048$ & $0.048$ &$-2.000$&$0$ & $-0.048$ & $-0.048$ &$-2.000$&$0.000$\\
\hline 
$0.100$ & $0.804$ & $-0.214$ &$0.532$&$4.935$ & $-0.804$ & $0.214$ &$0.522$&$4.955$\\
        & $-0.376$ & $0.480$ &$2.553$&$0.893$& $0.376$ & $-0.480$ &$2.553$&$0.893$\\
        & $0.066$ & $0.066$ &$-2.00$&$0.000$& $-0.066$ & $-0.066$ &$-2.00$&$0.000$\\
\hline 
$0.19$&$0.750$&$-0.210$&$0.560$ &$4.880$& $-0.750$& $0.210$& $0.565$&$4.880$ \\
      &$0.088$&$0.088$&$-2.00$ &$0.000$& $-0.088$& $-0.088$&$-2.000$&$0.000$\\
      &$0.402$&$-0.246$&$1.223$&$3.552$ & $-0.402$& $0.246$&$1.223$&$3.552$\\
\hline
$0.25$&$0.686$&$-0.206$&$0.600$&$4.799$ & $-0.686$&$0.206$&$0.600$&$4.799$\\
      &$0.100$&$0.100$&$-2.000$ &$0.000$&$-0.100$&$-0.100$&$-2.000$&$0.000$\\
      &$0.526$& $-0.208$&$0.791$&$4.418$&$-0.526$&$0.206$&$0.790$&$4.418$\\
\hline 
$0.6$&$0.142$&$0.142$& $-2.000$&$0.000$& $-0.142$&$-0.142$&$-2.000$&$0.000$\\
     &  $ $     &   $ $    &   $ $     &  $ $  &    $ $    &  $ $    & $ $       &$ $\\
\hline 
$1.0$&$0.173$&$0.173$& $-2.000$&$0.000$ & $-0.173$&$-0.173$&$-2.000$&$0.000$\\
     &  $ $  &   $ $ &    $ $  &  $ $   &     $ $ &   $ $  &  $ $   &$ $\\
\hline
\end{tabular}
\caption{Asymptotic solutions for $D=4$, $\theta=0.1$ with inclusion of the parameter $\bar{\lambda}$}
\label{table2}
\end{table}
\end{center}

%%%%%%%%%%%%%%%%%%%%%%%%%%%%%%%%%%%%%%%%%%%%%%%%%%%%%%%%%%%%%%%%%%%%%%%%%%%%%%%%%%%%%%%%%%%%%%%%%%%%%%%%%%%%%%%%%%%%%%%

\section{Analytical Solution for electromagnetic Field and Power Spectrum}
It is clear from the work of \cite{pahwa08} 
that within a few e-foldings, 
$a$ and $b$ enter an asymptotic regime,
where they become exponential functions of time.  
The exponents themselves are given in in Tables~(\ref{table1}) and Table~(\ref{table2}).
In such a case, one can 
obtain analytical solutions for the $\mathcal{A}(k,\eta)$.
It is convenient to express the evolution of the scale factors in terms of the conformal time ($\eta$) defined in equation \ref{etaref}.
In terms of conformal time, the evolution of the scale factors, given in Eq.~\ref{asssoln}, becomes
\begin{equation}
 a(\eta)=a_{0}\left|\frac{\eta}{\eta_{0}}\right|^{-1}, \hspace{2cm}  b(\eta)=b_{0}\left|\frac{\eta}{\eta_{0}}\right|^{-\beta/\alpha},
\end{equation}

By substituting this in equation ($25$), the function $V(\eta)$ takes the form
\begin{equation}
 V(\eta)=\frac{\xi(\xi-1)}{\eta^2}
 \label{pot}
\end{equation}
where,
\begin{equation}
\xi=\frac{D}{2}\left(\frac{-\beta}{\alpha}\right)
\label{zi}
\end{equation}
 Substituting Eq.~(\ref{pot}) in Eq.~(\ref{VP}) we get, 
\begin{equation}
\mathcal{A}''(k,\eta)+\left[k^2-\frac{\xi(\xi-1)}{\eta^{2}}\right]\mathcal{A}(k,\eta)=0
\end{equation}
whose solution can be obtained in terms of Hankel functions.
\begin{equation}
\mathcal{A}(k,\eta)=(-k\eta)^{1/2}[C_{1}(k)J_{\xi -1/2}(-k\eta)+C_{2}(k)J_{-\xi +1/2}(-k\eta)]
\end{equation}
where $C_{1}(k)$ and $C_{2}(k)$ are scale-dependent coefficients to be fixed by the initial conditions.
The length scales involved here are Hubble radius defined as $d_{H}=1/H$ and the physical length scale ($a/k$) associated with each mode. 
Since the evolution equation for the Electromagnetic field is a linear equation all modes will evolve independently. Hence, there will be different set
of initial conditions for different modes. 
Comparing the two length scales (i.e. $a/k$ and $d_{H}$), the modes will be said to be within the Hubble radius if $k/aH>1$ and outside the Hubble 
radius if $k/aH<1$. At horizon crossing, $k/aH=1$.
For exponential inflation $\eta=-1/aH$. This implies $k/aH \approx -k\eta$ and at horizon crossing, $-k\eta =1$. 
A given mode is therefore within the Hubble radius for $-k\eta >1$ and outside the Hubble  radius when $-k\eta < 1$. 
Substituting  the solution for the redefined variable $\mathcal{A}$ in Eq.(\ref{VP}), we deduce the form of 
power spectrum at super-horizon scales ($-k\eta<<1$) as \cite{my08,kandu10},
\begin{equation}
\frac{d\rho_{B}}{dln k}=\frac{\textit{F(n)}}{2\pi^2}H^4(\frac{k}{aH})^{4+2n}\approx \frac{\textit{F(n)}}{2\pi^2}H^4 (-k\eta)^{4+2n}
\label{PSB}.
\end{equation}
Here,
\begin{equation}
\textit{F}(n)=\frac{\pi}{2^{2n+1}\Gamma^{2}(n+\frac{1}{2})cos^{2}(\pi n)}.
\end{equation}
We have $n=\xi$ if $\xi \leq 1/2$ and $n=1-\xi$ if $\xi \geq 1/2$.
Similarly the power spectrum for the electric field is given by,
\begin{equation}
\frac{d\rho_{E}}{dln k}=\frac{\textit{G(m)}}{2\pi^2}H^4(\frac{k}{aH})^{4+2m}\approx \frac{\textit{G(m)}}{2\pi^2}H^4 (-k\eta)^{4+2m},
\end{equation}
where,
\begin{equation}
\textit{G}(m)=\frac{\pi}{2^{2m+1}\Gamma^{2}(m+\frac{1}{2})cos^{2}(\pi m)}.
\end{equation}
Here $m=1+\xi$ if $\xi \leq -1/2$ and $m=-\xi$ if $\xi \geq -1/2$.
The Hubble parameter remains almost constant during inflation. Hence the scale dependence of power spectrum comes only from the other
factor containing $k\eta$. It is evident that a scale invariant spectrum for magnetic field doesn't imply scale invariance of electric 
field power spectrum as $n\neq m$. A scale invariant power spectrum for magnetic field can be obtained for $n=-2$, 
which corresponds to two different values of $\xi$, namely $\xi=-2$ or $\xi=3$. We define 
the 
spectral index of the magnetic field as $n_{B}=4+2n$.
For the case $\xi =3$, the electric field spectrum grows rapidly with time 
and hence may lead to a strong back reaction \cite{my08,kandu10}. Hence, $\xi =3$ may not be a viable scenario. However, for $\xi = -2$ 
we do not have this problem of growing electric field while at the same time yielding a scale invariant 
magnetic power spectrum.  Hence the latter case is a better acceptable scenario for magnetic field generation. 
We have summarized in Table~\ref{table1} and Table~\ref{table2},
the values of $\xi$ and the magnetic spectral index $n_B=4+2n$ for different solutions obtained in section \ref{numerical}

For $\xi>0$ it is clear from Eq.~(\ref{zi}) that 
we require 
the scale factors of the normal and extra dimensions to 
have opposite behavior. In particular for the normal dimensional space to expand the higher dimensional space needs to contract. 
From the point of view of the evolution of universe, this is the qualitative situation we need. The case of $\xi=3$ falls in this category.
Similarly in order to have $\xi<0$ both the dimensions should have identical behavior. This is the case for $\xi=-2$. In 
this situation we need to have either both normal and higher dimensional space to expand or for both to contract. The situation in which normal
space contracts is not acceptable. The other situation in which normal
dimensions expand has a problem that the higher dimensional space too expands. However, it may be possible to construct a 
mechanism which can freeze the expansion of the higher dimensional space at a suitable scale 
(or even make it recollapse), 
to avoid conflicts with observations.

We examine the behavior of magnetic field spectrum for the solutions obtained in Table~(\ref{table1}) and Table~(\ref{table2}).
To begin with one needs to restrict oneself to stable solutions.
In Table~(\ref{table1}), for $D=3$ we have stable solutions for both $n_{B}>0$ and $n_{B}<0$ ($n_{B}=0$ refers to scale invariant case).
For the cases where $n_{B}=4.854$ we have expanding normal dimensional space with contacting higher dimensional space ( for
example, $\theta=0.1$, $\alpha=0.809$, $\beta=-0.309$). Although the feature of expansion is acceptable, the value of $n_{B}$
indicates that the spectrum of magnetic field produced will be strongly blue. One can have a vastly growing magnetic field spectrum for
$n_{B}=-1.854$. This, however, can cause severe back-reaction problem. Also in this case the normal dimensional space contracts while 
the higher dimensional space expands (for example, $\theta=0.1$, $\alpha=-0.309$, $\beta=0.809$),  
Even if we construct a mechanism to halt the expansion of higher dimensions, the fact that normal dimensions are contacting makes 
it an unacceptable scenario. For $D=2$, independent of the value of $\theta$, we get $n_{B}=4.556$.
This is again a highly blue spectrum. 
For cases $D=4$ and $D=10$ we have $n_{B}>0$, independent of which of the dimensions
are expanding and which are contacting. In general, the cases where normal dimensions expand while higher dimensions contract 
the values $n_{B}$ are very different from $n_{B}=0$ as compared to the reverse cases where normal dimensions contract and 
higher dimensions expand.
There are however some interesting features in Table~(\ref{table2}) which is specifically for the case of $D=4$ and with a
cosmological constant parameter ($\bar{\lambda}$). We again focus only on the stable solutions, where also the normal dimensions are expanding.
There are then two types of solutions. The first kind is very similar to the vacuum case,
and obtained for a small enough $\bar{\lambda}$. Here, we again get a blue spectrum (for example, $\bar{\lambda}=0.001$; $\alpha=0.846$, 
$\beta=-0.218$, $n_{B}=4.969$ \& 
$\bar{\lambda}=0.25$; $\alpha=0.686$, $\beta=-0.206$, $n_{B}=4.799)$). 
More interestingly, for $\bar{\lambda}\neq 0$ we do have cases where $n_{B}=0$ (i.e. perfect scale invariant case). 
We also note that such cases have stable solutions with both the normal as well as higher dimensions expand simultaneously and at the same 
rate ( for eg., $\bar{\lambda}=0.001$; $\alpha=0.007$, $\beta=0.007$, $\bar{\lambda}=0.050$; $\alpha=0.048$, 
$\beta=0.048$). There exist such solution for all non-zero $\bar{\lambda}$. 
In fact, for these solutions with $\alpha=\beta$, $\psi = -D/2$ and $n_B = 0$ exactly for $D=4$.
Since the values of exponents are dependent upon $\bar{\lambda}$, the parameter can be tuned to get admissible values of these exponents. 
However, in this case the mechanism(s) to freeze the expansion of higher dimensional space is needed. 
Alternatively, one needs a model where an effective $\bar{\lambda}$ can be turned off to
join onto a vacuum type solution, with expanding normal dimension and contracting extra dimensions.
These models would have to perhaps invoke a higher dimensional scalar field with a potential.
Discussions about the details of such a mechanism is beyond the scope of this paper.

%%%%%%%%%%%%%%%%%%%%%%%%%%%%%%%%%%%%%%%%%%%%%%%%%%%%%%%%%%%%%%%%%%%%%%%%%%%%%%%%%%%%%%%%%%%%%%%%%%

\section{Magnetic Field Strength}
From Eq.(\ref{PSB}), for a perfect scale invariant spectrum for Magnetic field, we have, ($n_{B}=0$)
\begin{equation}
\frac{d\rho_{B}}{dln k}=\frac{\textit{F(n=-2)}}{2\pi^2}H^4=\frac{\textit{9}}{4\pi^2}H_{f}^4
\label{PSBin}.
\end{equation}
where, $H_{f}$ is the value of Hubble parameter which remains constant during inflation. 
Having identified the parameters for obtaining both a scale invariant magnetic field spectrum and at the same time, 
which does not lead to a growing electric filed power spectrum, we now estimate the strength of the magnetic field. In our case the role 
of extra dimension is present only before the end of inflation, because the extra dimensions are assumed to be frozen 
afterwards. Thus in the post inflationary era, the energy density in the magnetic field evolves as,
\begin{equation}
 \rho_{B}(0)=\rho_{B}(f)\left(\frac{a_{f}}{a_{0}}\right)^4
\end{equation}
where, $\rho_{B}(0)$ \& $\rho_{B}(f)$ are the magnetic field energy densities and $a_{0}$ \& $a_{f}$ are the 
values of scale factor at present epoch and at the end of inflation respectively. The ratio of the present 
value of scale factor to that at the end of inflation depends upon the history of universe. We assume that the 
universe entered into reheating phase almost instantly after the end of inflation.
The entropy of the universe is constant through its evolutions i.e. $ga^3T^3$ $\approx$ $const$, where $g$ represents the relativistic
degrees of freedom at a particular epoch while $T$ is the temperature of the fluid at that epoch. Using the 
entropy conservation we get,
\begin{equation}
 \frac{a_{f}}{a_{0}}\approx\frac{g_{0}^{1/3}}{g_{f}^{1/12}}\frac{T_{0}}{H_{f}^{1/2}Mpl^{1/2}}\left(\frac{8\pi^3}{90}\right)^{1/4}.
 \label{scale}
\end{equation}
Here, $g_{f}$ and $g_{0}$ are the relativistic degrees of freedom at the end of inflation and current epoch respectively.
Taking $g_{f} \sim 100$, Eq.~(\ref{scale}), gives $a_{0}/a_{f}\sim 5\times 10^{31}(H_{f}/Mpl)^{1/2}$. For a viable cosmology $b(t)$ should saturate
at an admissible value. The possible mechanism which may help in this is beyond the scope of this paper and needs to be explored further.
For the extra dimensions frozen at the scale of higher dimensional Planck mass, we can write $M \sim Mpl$. 
Therefore we assume $H_{f}=\kappa Mpl$, where $\kappa$ sets the scale of inflation. Depending on the numerical value of $\kappa$, 
it is possible to generate magnetic fields which can have a strength today of the order of a nano gauss.
From Table($2$) depending on the value of the parameters, magnetic fields of different strengths can be generated. 
By changing the value of $\bar{\lambda}$ we can have different values of $\alpha$ (which is in turn related to $\kappa$ as the scale of inflation) for 
scale invariant magnetic fields ($n_B=0$). 
From Table~(\ref{table2}), for $\bar{\lambda}=0.001$, the value of $\kappa$ is of the order of $10^{-3}$. For this value of $\kappa$ we can generate magnetic
fields which at present are of order $10^{-9}G$.
\footnotetext[2]{Note that for all the other permissible non-scale invariant cases, $n_B$ is so large that there will be negligible magnetic fields 
on Mpc scales of interest.}
%%%%%%%%%%%%%%%%%%%%%%%%%%%%%%%%%%%%%%%%%%%%%%%%%%%%%%%%%%%%%%%%%%%%%%%%%%%%%%%%%%%%%%%%%%%%%%%%%%%%%%%%%%%%%%

\section{Conclusions}
In this paper, an attempt has been made to investigate the possibility of generating primordial cosmic magnetic fields in certain cosmological
models motivated by higher dimensions. The approach in 
this paper is to include the Gauss-Bonnet term in the action. The evolution of the scale factors of the normal as well as the extra dimensions
is governed by the Gauss-Bonnet term and the number of extra dimensions. We found that conformal invariance of the Electromagnetic action
is naturally broken by an evolving scale factor of the extra dimensions. Within a few e-foldings the scale factors in both
normal dimension and the extra dimension assume an exponential form which makes it easier to obtain  analytical solutions for the vector potential.
A scale invariant magnetic field spectrum requires $4+2n=0$, which corresponds to $\xi=3$ or $\xi=-2$.

The power spectrum corresponding to $\xi=3$ has the problem of vastly growing electric fields. As shown in 
Table~(\ref{table1}) \& Table~(\ref{table2}). We do not have stable solutions which can give a nearly scale invariant spectrum 
without including parameter $\bar{\lambda}$. 
The cases for which lower $n_{B}$ values are obtained are not admissible as it requires deflationary solutions for normal dimensions. 
Power spectrum corresponding to  $\xi=-2$ doesn't suffer from the problem of growing electric fields. 
This case is, however, only possible when normal as well as extra dimensions are both growing and contracting simultaneously.
By including the parameter $\bar{\lambda}$ we have obtained such solutions for $D=4$ in Table~(\ref{table2}). A perfect scale invariant 
spectrum for magnetic field is naturally
obtained as $n_{B}=0$. The parameters $\bar{\lambda}$ and $\chi$ can be tuned to obtain significant 
amount of magnetic field. A suitable mechanism to freeze the expanding extra dimensions is needed and 
we hope to explore such possibilities further.
%%%%%%%%%%%%%%%%%%%%%%%%%%%%%%%%%%%%%%%%%%%%%%%%%%%%%%%%%%%%%%%%%%%%%%%%%%%%%%%%%%%%%%%%%%%%%%%%%%%%%%%%%%%%%%%%%%%%%%%%%%%%%%%%%%%%%%%%%%

\acknowledgments{KA, IP and TRS acknowledge the facilities at the IUCAA Resource Centre, University of Delhi.  
IP acknowledges the CSIR, India for
assistance under grant 09/045(0908)/2009-EMR-I. KA acknowledges the UGC, India for
assistance under grant AA/139/F-42/2009-10. TRS acknowledges the CSIR, India for
assistance under grant O3(1187)/11/EMR-II.}

\bibliographystyle{hunsrt}
\bibliography{akis}

\end{document}